\documentclass{article}

\PassOptionsToPackage{numbers, compress}{natbib}

\usepackage[final]{tackling_climate_workshop_style}

\usepackage[utf8]{inputenc} 
\usepackage[T1]{fontenc}    
\usepackage{hyperref}       
\usepackage{url}            
\usepackage{booktabs}       
\usepackage{amsmath} 
\usepackage{amsfonts}       
\usepackage{nicefrac}       
\usepackage{microtype}      
\usepackage{makecell}
\usepackage{threeparttable}
\usepackage{graphicx}
\usepackage{algorithm}
\usepackage{algpseudocode}
\usepackage{dsfont}

\title{Multi-Agent Reinforcement Learning for Global Climate Policy Cooperation}

\author{%
  Anne Moll-Elsborg\\
  Technical University of \\ Denmark\\
  Department of Technology, \\ Management and Economics\\
  \texttt{anne@mollelsborg.com} \\
  \And
  Oskar Bohn Lassen \\
  Technical University of \\ Denmark \\
  Department of Technology, \\ Management and Economics \\
  \texttt{obola@dtu.dk} \\
  \And
  Francisco Camara Pereira \\
  Technical University of \\ Denmark \\
  Department of Technology, \\ Management and Economics \\
  \texttt{camara@dtu.dk} \\
}

\begin{document}

\maketitle

\begin{abstract}
International climate policy requires coordination among countries whose mitigation decisions impose local costs but generate global benefits.
Classical game-theoretic and integrated assessment approaches can study these incentives, but typically rely on prescribed behavioural assumptions or equilibrium concepts. 
Recent work has therefore explored Multi-Agent Reinforcement Learning as a way to model climate-policy cooperation as an emergent outcome of repeated strategic interaction.
We extend an existing MARL climate framework with a multi-stage negotiation mechanism in which agents propose agreements, accept or reject commitments, and face penalties for non-compliance. 
This allows us to study how agreement design affects learned mitigation behaviour, compliance, and climate outcomes.
In an idealized setting with four homogeneous agents, we find that non-binding agreements do little to overcome free-riding, while moderate non-compliance penalties improve cooperation and reduce warming.
However, excessively strong penalties discourage ambitious commitments, revealing a trade-off between enforcement and participation.
We position the framework as a methodological contribution for testing agreement design rather than a predictive model of real negotiations.
\end{abstract}

\section{Introduction}

Climate change is a global cooperation problem: Emissions affect the climate regardless of their source, while mitigation costs are borne individually and their benefits are shared through reduced climate damages. Countries may therefore under-invest, free-ride, or defect from agreements, despite the long-term risks of delayed mitigation~\cite{ipcc2021_wg1, ipcc2022_wg2}. International policy addresses these incentives through agreements, commitments, and economic instruments, but their effects are difficult to study as climate policy involves repeated interaction, delayed outcomes, and strategic responses.

Integrated assessment models (IAMs) and game-theoretic models are widely used to analyse mitigation pathways, welfare, and strategic incentives~\cite{vanBeek2020_anticipating_futures_iam}. However, they often rely on assumed behaviour, foresight, and equilibrium concepts~\cite{madani2013_modeling_climate_negotiations}. 
A repeated public-goods game with exogenous sanctions captures the basic free-rider logic, but becomes too complex to solve once commitments form over time, actions are multi-dimensional, and payoffs depend on a nonlinear, delayed emissions-to-temperature response or on local climate-damage.
Multi-Agent Reinforcement Learning (MARL) complements these approaches by allowing agents to learn climate-policy strategies through repeated interaction, so cooperation, free-riding, and responses to agreement mechanisms emerge from learning rather than from an imposed equilibrium concept~\cite{ruddjones2025multi}.

Concrete MARL environments for climate cooperation have so far captured different parts of the problem.
\citet{RICEN} model regions that negotiate agreements before choosing climate and economic actions, building on the RICE model~\cite{nordhaus1996rice} to represent regional economies, emissions, and climate damages. 
\citet{CICERO} instead use a neural surrogate of the CICERO Simple Climate Model (CICERO-SCM)~\cite{sandstad2024cicero} to make a detailed emissions-to-temperature response feasible during MARL training.
However, their framework does not include negotiation or agreement-formation actions and cooperation therefore depends solely on coordination emerging through learning. 
Experiments using their tool show that this is insufficient: Agents bear their own mitigation costs while sharing the benefits of reduced warming, leading learned policies to under-invest relative to a coordinated naive benchmark.
Existing frameworks thus offer either negotiation or a detailed climate response, but not both.

We close this gap by extending the environment of \citet{CICERO} with a multi-stage negotiation mechanism in which agents learn which commitments to propose, accept, and honour. 
We deliberately keep the experiment simple, with homogeneous agents, so that heterogeneity does not blur the results of this proof of concept.

\section{Methodology}
\label{sec:framework}

\begin{figure}[t]
  \centering
  \includegraphics[width=\linewidth]{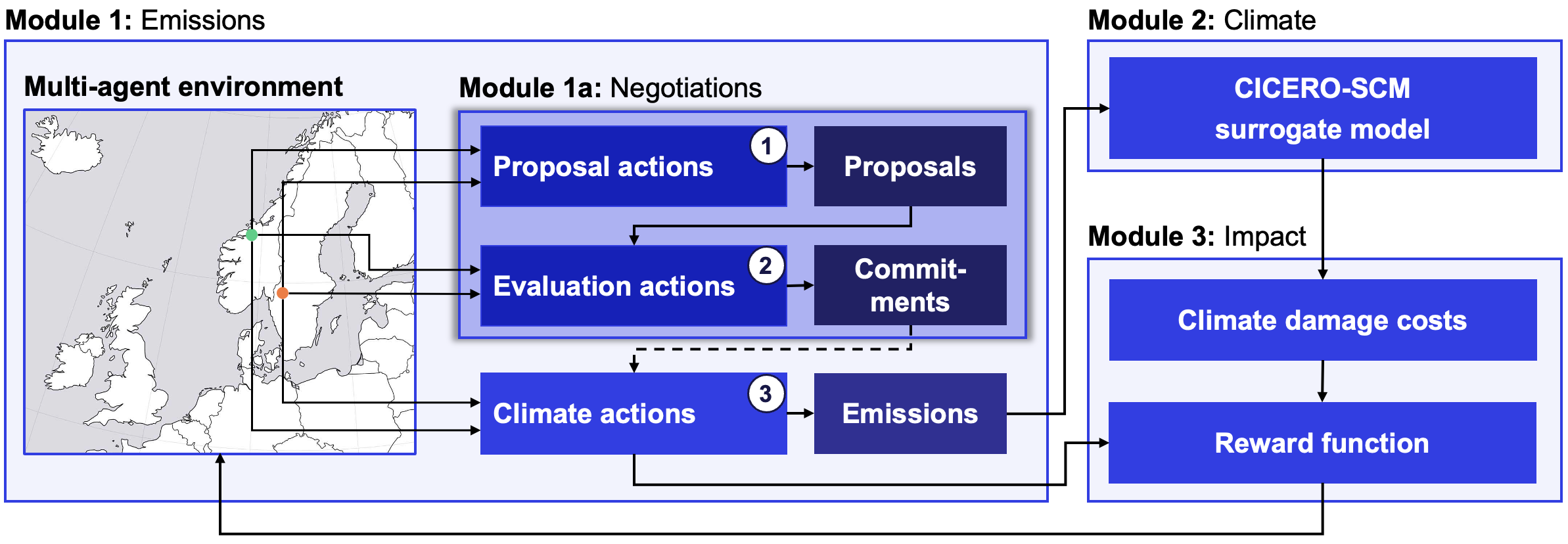}
  \caption{Multi-stage MARL framework. Module 1a: Agents propose commitments \textit{(stage 1)} and evaluate them \textit{(stage 2)}. Module 1: They choose climate actions \textit{(stage 3)}, which determine emissions. Module 2: Temperature change is predicted with the CICERO-SCM surrogate. Module 3: Temperature change is converted into climate-damage costs and rewards.}
  \label{fig:framework}
\end{figure}

In the environment of \citet{CICERO}, country-level climate actions determine emissions, temperature change, and agent-specific costs through a neural surrogate of CICERO-SCM \cite{sandstad2024cicero}. 
Our negotiation module separates agreement formation from climate-action implementation (Figure~\ref{fig:framework}).

\subsection{Environment.}

The environment is a finite-horizon multi-agent climate game with agents $\mathcal{N} = \{1, \dots, N\}$, each representing an idealized country or region, played annually from 2016 to 2085 ($H=70$ policy years). In each policy year $t$, agent $i$ chooses its climate actions, 
\begin{equation}
a^{\mathrm{clim}}_{it}
=
(m_{it},p_{it})
\in
\mathcal{M}\times\mathcal{P},
\end{equation}
where 
\begin{equation}
m_{it}\in\mathcal{M}=\{0,0.5,1\}^3,
\qquad
p_{it}\in\mathcal{P}=\{0,0.03,0.08\}.
\end{equation}

Here, $m_{it}$ sets the effort on three mitigation levers (energy decarbonisation, methane abatement, and land use) and $p_{it}$ is the fraction of output invested in preventive measures. Preventive investments accumulates into a decaying preventive stock $P_{it}\in[0,P_{\max}]$. 
Mitigation reduces each agent's emission growth relative to a baseline.
The agents' joint mitigation actions determine aggregate emissions $E_t$, which the CICERO-SCM surrogate model translates into global temperature change $\Delta T_t$. We denote the climate damage experienced by agent $i$ at time $t$ by $D_{it}$, which increases with temperature change and is reduced by accumulated preventive investment:

\begin{equation}
    D_{it} \propto \Delta T_t ^4 \times (1-P_{it}).
\end{equation}

The policy cost, $C_{it}^{\mathrm{pol}}$, is quadratic in mitigation effort and linear in preventive investment. At the end of each policy year, each agent receives a reward, the negative sum of climate-damage and policy costs:

\begin{equation}
r_{it} \propto -\left(D_{it} + C_{it}^{\mathrm{pol}}\right).
\label{eq:reward}
\end{equation}

A 15-year look-ahead of climate damages is added to the terminal reward. In our experiments, all agents have identical cost and damage coefficients. Agents observe a common global state: temperature, the year, each agent's emissions and preventive stock, and, in the multi-stage setting, the current stage and all proposals, acceptances, and commitments.

\begin{algorithm}[t]
\caption{Sequential multi-stage decision cycle}
\label{alg:mscycle}
\footnotesize

\begin{algorithmic}[1]

\Statex \textbf{Notation:}
\Statex $t\in\{1,\ldots,H\}$: policy year in an episode of length $H$.
\Statex $\sigma\in\{1,2,3\}$: decision stage (proposal, evaluation, climate).
\Statex $i\in\{1,\ldots,N\}$: agent index.

\Statex $E_t$: aggregate emissions.
\Statex $\Delta T_t$: global temperature change relative to 1900.
\Statex $f_{\theta_i}(\cdot)$: policy network of agent $i$.
\Statex $a_{it}
=
\left(
a_{it}^{\mathrm{prop}},
a_{it}^{\mathrm{eval}},
a_{it}^{\mathrm{clim}}
\right)
=
f_{\theta_i}(S_t)$:
full action vector of agent $i$.

\Statex $S_t=(E_t,\Delta T_t,a_{it})$: global environment state.

\Statex $a_{it}^{\sigma}=\mathtt{Mask}_{\sigma}(a_{it})$:
selects a component of $a_{it}$
where
$a_{it}^{(1)}=a_{it}^{\mathrm{prop}}$,
$a_{it}^{(2)}=a_{it}^{\mathrm{eval}}$, and
$a_{it}^{(3)}=a_{it}^{\mathrm{clim}}$.

\Statex $\mathcal{F}_{\sigma}(\cdot)$: updates the global environment state .

\Require Initial state $S_1$

\For{$t=1,\ldots,H$}
    \For{$\sigma=1,2,3$}
        \For{$i=1,\ldots,N$}
            \State $a_{it} \gets f_{\theta_i}(S_t)$
            \State $a_{it}^{(\sigma)}
            \gets \mathtt{Mask}_{\sigma}(a_{it})$
        \EndFor

        \State $S_t \gets
        \mathcal{F}_{\sigma}
        \left(
        S_t,
        (a_{it}^{(\sigma)})_{i=1}^{N}
        \right)$
    \EndFor
    \State $S_{t+1} \gets S_t$
\EndFor

\end{algorithmic}
\end{algorithm}

\subsection{Multi-stage decision cycle.}

We implement a sequential negotiation protocol, the core extension of this paper, inspired by RICE-N~\cite{RICEN}. 
It separates agreement formation from climate-policy implementation (Module~1a, Figure~\ref{fig:framework}). 
Each policy year comprises three steps and visits three stages:
\begin{equation}
\sigma_t \in \{1,2,3\},
\qquad
\sigma_t =
\begin{cases}
1, & \text{proposal},\\
2, & \text{evaluation},\\
3, & \text{climate}.
\end{cases}
\end{equation}
At $\sigma_t=1$, agents propose agreement terms:
\begin{equation}
a^{\mathrm{prop}}_{it}
\in
\left(\mathcal{A}^{\mathrm{prop}}\right)^{N-1},
\qquad
\mathcal{A}^{\mathrm{prop}}
=
\{0,0.5,1\}^2,
\end{equation}
where $\mathcal{A}^{\mathrm{prop}}$ is the discrete proposal space.
At $\sigma_t=2$, agents accept or reject proposals through evaluation actions:
\begin{equation}
a^{\mathrm{eval}}_{it}
\in
\{0,1\}^{N-1},
\end{equation}
determining active agreements. Negotiation stages yield zero immediate reward; their effects arise only in the climate stage. 
At $\sigma_t=3$, agents choose climate actions, $a^{\mathrm{clim}}_{it}$, the climate system advances by one year, and a reward is issued according to Equation~\ref{eq:reward}. 
Each agent has an independent policy, $f_{\theta_i}$, implemented as an LSTM actor-critic and trained with Proximal Policy Optimization (PPO)~\cite{schulman2017ppo} to maximise its own expected discounted return with discount factor $\gamma=0.999$, using the reward in Equation~\ref{eq:reward} as the learning signal. 
Algorithm~\ref{alg:mscycle} summarises the sequential decision cycle.

\subsection{Agreement mechanism.} 
We implement bilateral agreements, in which each proposal action ($a^{\mathrm{prop}}_{it}$) specifies, a \textit{promise} and a \textit{request} of the effort it asks of that agent, each taking a value in $\{0,0.5,1\}$.
Agent $i$ faces a penalty $c^{\mathrm{agr}}\geq0$ when its implemented effort $m_{it}$ on the negotiated lever, methane abatement, falls below the commitment with the largest effort, $\bar m_{it}$.
Hence, if multiple accepted agreements apply to an agent, its effective commitment is the highest accepted mitigation level.
The agreement-related cost $C^{\mathrm{agr}}_{it}$ is introduced to the reward in Equation~\ref{eq:reward}:
\begin{equation}
r_{it} \propto -\left(D_{it} + C_{it}^{\mathrm{pol}} + C_{it}^{\mathrm{agr}}\right),
\quad C_{it}^{\mathrm{agr}} = \mathds{1}\!\left[m_{it} < \bar m_{it}\right] c^{\mathrm{agr}}
\label{eq:ms_reward}
\end{equation}
where the indicator equals one when agent $i$ falls short of its commitment and zero otherwise. Non-binding agreements ($c^{\mathrm{agr}}=0$) test whether agents can learn to collaborate without penalties for non-compliance. 

\subsection{Benchmark policies and metrics.}

Learned policies are evaluated against two fixed policies. Under the \textit{equal-effort} benchmark, every agent applies identical mitigation actions and preventive investments in every policy year, providing a coordinated reference point. The \textit{unilateral-deviation} benchmark is identical except that one agent's mitigation actions are reduced, revealing the private gain from defecting while all other agents maintain equal effort.

\section{Results}

The experiments examine how agreement mechanisms shape learned behaviour in the multi-stage MARL environment, and how these behavioural responses affect mitigation, compliance, and climate outcomes. Homogeneous agents make equal effort a coordinated benchmark, so deviations reflect strategy rather than country differences.

\begin{table}[t]
\centering
\setlength{\tabcolsep}{2pt}
\renewcommand{\arraystretch}{1.1}
\caption{Four homogeneous agents comparing final greedy policies to the fixed equal-effort and unilateral-deviation policies. Showing individual reward, collective reward, mitigation share, and peak temperature change.}
\label{tab:nonbinding}
\begin{threeparttable}
\scriptsize
\begin{tabular}{llcccccccccccc}
\toprule
& & \multicolumn{4}{c}{\textbf{Final greedy policy range}}
& \multicolumn{4}{c}{\textbf{Equal effort}}
& \multicolumn{4}{c}{\textbf{Unilateral deviation}} \\
\cmidrule(lr){3-6} \cmidrule(lr){7-10} \cmidrule(lr){11-14}
\makecell{\textbf{Climate}\\\textbf{damage}} & \makecell{\textbf{Penalty}}
& \makecell{\textbf{Ind.}\\\textbf{reward}} & \makecell{\textbf{Col.}\\\textbf{reward}}
& \makecell{\textbf{Mit.}\\\textbf{share}} & \makecell{$\Delta T$\\\textbf{peak}}
& \makecell{\textbf{Ind.}\\\textbf{reward}} & \makecell{\textbf{Col.}\\\textbf{reward}}
& \makecell{\textbf{Mit.}\\\textbf{share}} & \makecell{$\Delta T$\\\textbf{peak}}
& \makecell{\textbf{Ind.}\\\textbf{reward}} & \makecell{\textbf{Col.}\\\textbf{reward}}
& \makecell{\textbf{Mit.}\\\textbf{share}} & \makecell{$\Delta T$\\\textbf{peak}} \\
\midrule
Low & None
& \makecell{[$-$12.70;\\ $-$12.38]} & \makecell{[$-$50.15;\\ $-$50.12]}
& \makecell{[0.01;\\ 0.59]} & \makecell{[1.566;\\ 1.570]}
& $-$12.40 & $-$49.60 & 0.25 & 1.512
& \makecell{$-$12.27;\\ $-$12.45} & $-$49.62 & \makecell{0.18;\\ 0.27} & 1.524 \\
\addlinespace
Medium & None
& \makecell{[$-$23.00;\\ $-$21.94]} & \makecell{[$-$89.50;\\ $-$88.15]}
& \makecell{[0.13;\\ 0.46]} & \makecell{[1.539;\\ 1.560]}
& $-$21.78 & $-$87.12 & 0.25 & 1.512
& \makecell{$-$21.69;\\ $-$21.87} & $-$87.32 & \makecell{0.18;\\ 0.27} & 1.524 \\
\addlinespace
High & None
& \makecell{[$-$54.07;\\ $-$50.35]} & \makecell{[$-$210.33;\\ $-$206.01]}
& \makecell{[0.12;\\ 0.44]} & \makecell{[1.488;\\ 1.534]}
& $-$47.71 & $-$190.83 & 0.25 & 1.442
& \makecell{$-$47.42;\\ $-$48.12} & $-$191.76 & \makecell{0.18;\\ 0.27} & 1.446 \\
\bottomrule
\end{tabular}
\begin{tablenotes}[flushleft]\scriptsize
\item \textit{Note.} Final greedy policy reports the range across three runs under non-binding bilateral agreements, where agents can propose and evaluate agreements but are not forced to comply. The unilateral-deviation policy is the equal-effort policy with country~0's mitigation effort reduced; its individual reward and mitigation share are reported as the deviating agent above the three others.
\end{tablenotes}
\end{threeparttable}
\end{table}

Under \emph{non-binding} bilateral agreements, agents can propose and accept commitments, but commitments do not restrict actions or change payoffs. Table~\ref{tab:nonbinding} shows that, across the three damage levels, the equal-effort benchmark outperforms learned policies and mitigation remains uneven. The unilateral-deviation test explains this result: Reducing one agent's mitigation effort increases that agent's reward, but lowers the rewards of the other agents and collective welfare. Higher damages narrow the dispersion in mitigation shares but do not remove this free-rider incentive. Under \emph{binding} agreements, agents that implement effort below an accepted commitment receive a non-compliance penalty. Table~\ref{tab:binding} shows that, in the medium-damage setting, a low penalty improves collective reward from $[-89.50; -88.15]$ to $[-85.86; -85.10]$, above the equal-effort benchmark of $-87.12$, and narrows mitigation shares to $[0.19; 0.33]$; however, 5--15 of 69 commitments are violated. A medium penalty eliminates violations while maintaining similar rewards and accepted mitigation up to $0.83$. With a high penalty, agents avoid ambitious commitments: Accepted mitigation falls to at most $0.32$ and collective reward declines to $[-88.53; -85.71]$. Penalties therefore affect both compliance and participation through learned incentives.

\subsection{Impact.}
The framework makes climate-agreement design experimentally testable within a scalable MARL environment that preserves an emissions-to-temperature response. In a deliberately simple setting, it reproduces the central enforcement trade-off: Stronger penalties can increase compliance and mitigation, but may also discourage participation and ambition. By letting these effects emerge from learned behaviour, the approach provides a useful framework for assessing how countries may respond to agreement mechanisms before they are adopted.

\subsection{Limitations and scope.}
\begin{table}[t]
\centering
\scriptsize
\setlength{\tabcolsep}{3pt}
\renewcommand{\arraystretch}{1.1}
\caption{Four homogeneous agents under medium climate damage costs and binding bilateral agreements with different non-compliance penalty levels. Showing individual/collective climate reward, individual/collective non-compliance penalty, mitigation share, peak temperature change, and agreement-related measures: mean accepted mitigation level per agent, number of accepted proposals per agent, and total accepted mitigation effort per agent.}
\label{tab:exp2b_binding_bilateral}

\begin{threeparttable}
\begin{tabular}{llccccccccc}
\toprule
& & \multicolumn{9}{c}{\textbf{Final greedy policy range}} \\
\cmidrule(lr){3-11}
\makecell{\textbf{Climate}\\\textbf{damage}}
& \makecell{\textbf{Penalty}}
& \makecell{\textbf{Ind. climate}\\\textbf{reward}}
& \makecell{\textbf{Col. climate}\\\textbf{reward}}
& \makecell{\textbf{Ind.}\\\textbf{penalty}}
& \makecell{\textbf{Col.}\\\textbf{penalty}}
& \makecell{\textbf{Mit.}\\\textbf{share}}
& \makecell{$\Delta T$\\\textbf{peak}}
& \makecell{\textbf{Mean accepted}\\\textbf{mit. level}}
& \makecell{\textbf{Accepted}\\\textbf{proposals}}
& \makecell{\textbf{Accepted}\\\textbf{mit. effort}} \\
\midrule
Medium & None
& \makecell{[$-$23.00;\\ $-$21.94]}
& \makecell{[$-$89.50;\\ $-$88.15]}
& \makecell{-}
& \makecell{-}
& \makecell{[0.13;\\ 0.46]}
& \makecell{[1.539;\\ 1.560]}
& \makecell{-}
& \makecell{-}
& \makecell{-} \\
\addlinespace
Medium & Low
& \makecell{[$-$21.56;\\ $-$21.18]}
& \makecell{[$-$85.86;\\ $-$85.10]}
& \makecell{[$-$0.08;\\ 0.00]}
& \makecell{[$-$0.15;\\ $-$0.05]}
& \makecell{[0.19;\\ 0.33]}
& \makecell{[1.470;\\ 1.483]}
& \makecell{[0.11;\\ 0.68]}
& \makecell{[21;\\ 121]}
& \makecell{[4.18;\\ 63.92]} \\
\addlinespace
Medium & Medium
& \makecell{[$-$21.92;\\ $-$21.11]}
& \makecell{[$-$85.57;\\ $-$85.25]}
& 0.00
& 0.00
& \makecell{[0.18;\\ 0.30]}
& \makecell{[1.467;\\ 1.473]}
& \makecell{[0.19;\\ 0.83]}
& \makecell{[3;\\ 138]}
& \makecell{[1.50;\\ 53.00]} \\
\addlinespace
Medium & High
& \makecell{[$-$23.72;\\ $-$21.28]}
& \makecell{[$-$88.53;\\ $-$85.71]}
& 0.00
& 0.00
& \makecell{[0.17;\\ 0.33]}
& \makecell{[1.467;\\ 1.476]}
& \makecell{[0.00;\\ 0.32]}
& \makecell{[0;\\ 117]}
& \makecell{[0.00;\\ 35.10]} \\
\bottomrule
\end{tabular}

\begin{tablenotes}[flushleft]\scriptsize
\item \textit{Note.} Final greedy policy reports the range across three runs. Climate reward includes climate damages from temperature change and the costs of mitigation and preventive investment. The non-compliance penalty is applied when agents accept an agreement but do not follow through.
\end{tablenotes}
\end{threeparttable}
\label{tab:binding}
\end{table}
The experiment uses four homogeneous agents with identical costs and damages which isolates strategic behaviour but removes the asymmetries that dominate real negotiations such as income, individual damages, mitigation opportunities, and global bargaining power. 
Agreements are bilateral promise/request pairs, whereas most real agreements, notably the Paris Agreement, are multilateral pledge-and-review systems that rely on transparency, reputation, and reciprocity rather than central sanctions~\cite{falkner2016paris, keohane2016cooperation}.
The framework should be seen as a proof of concept, not real policy evaluation, and the results therefore support qualitative statements about mechanisms, not quantitative predictions about countries.

\begin{ack}
This work is supported by Novo Nordisk Foundation grant NNF23OC0085356 and Pioneer Centre for AI, DNRF grant number P1.
\end{ack}

\bibliographystyle{ACM-Reference-Format} 
\bibliography{references}

\end{document}